\documentclass[aps,prb,showpacs,twocolumn]{revtex4-1}
\usepackage{amssymb}
\usepackage{amsmath}
\usepackage{graphicx}
\usepackage{epsfig}

\begin{document}

\title{Moir\'{e} pattern of spin liquid and Neel magnet in a Kitaev chain}
\author{R. Wang, P. Wang, K. L. Zhang, and Z. Song}
\email{songtc@nankai.edu.cn}
\affiliation{School of Physics, Nankai University, Tianjin 300071, China}
\begin{abstract}
A moir\'{e} pattern occurs when two periodic structures in a system have a
slight mismatch period, resulting the coexistence of distinct phases in
different large-scale spacial regions of the same system. Two periodic
structures can arise from periodic electric and magnetic fields,
respectively. We investigated the moir\'{e} pattern via a dimerized Kitaev spin
chain with a periodic transverse field, which can be mapped onto the system
of dimerized spinless fermions with p-wave superconductivity. The exact
solution for staggered field demonstrated that the ground state has two
distinct phases: (i) Neel magnetic phase for nonzero field, (ii) Spin liquid
phase due to the emergence of isolated flat Bogoliubov--de Gennes band for
vanishing field. We computed the staggered magnetization and local density
of states (\textrm{LDOS}) for the field with a slight difference period to
the chain lattice. Numerical simulation demonstrated that such two phases
appear alternatively along the chain with a long beat period. Additionally,
we proposed a dynamic scheme to detect the Moir\'{e} fringes based on the
measurement of Loschmidt echo (\textrm{LE}) in the presence of local
perturbation.
\end{abstract}

\maketitle

\section{Introduction}

The moir\'{e} patterns emerge due to the\ superposition of two periodic
structures, with either slightly different period or different orientations,
and have been realized in materials \cite%
{M.Yan,L.A.,C.R.,B.Hunt,C.Woods,YT,Bloch}. Recently there has been a growing
interest in the influence of the moir\'{e} pattern in physical systems. The
moir\'{e} pattern as a new way to apply periodic potentials in van der Waals
heterostructures to tune electronic properties, has been extensively studied
\cite{L.A.,C.R.,B.Hunt,Gorbachev,Song,Jung}. Many interesting phenomena have
been observed in the heterostructure\ materials with small twist angles and
mismatched lattice constants. moir\'{e} patterns in condensed matter systems
are produced by the difference in lattice constants or orientation of two
two-dimensional lattices when they are stacked into a two-layer structure.

\begin{figure}[tbph]
\includegraphics[ bb=16 320 455 650, width=0.35\textwidth, clip]{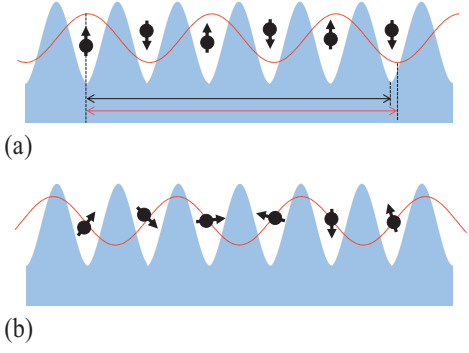}
\caption{Schematic illustration of two distinct phases is induced by a
slight mismatch of periodic electric field (blue area) and magnetic field
(red line). The electric field acts as a periodic potential (realized by
optical lattice) confines the particles as an array, while the magnetic
field affects the spin orientation of the particles. The period difference
is indicated by two arrowed lines (black and red), which results in two
typical configurations of two fields in large scale: (a) The staggered field
get its maximal magnitude, while (b) vanishes. Two distinct phases are
formed as (a) Neel magnetic phase, (b) spin liquid phase due to the isolated
flat band (see the text a complete description).}
\label{fig1}
\end{figure}

While most studies of this field have focused on the quasi-two-dimensional
system, the one-dimensional moir\'{e} system is by far less well
investigated and is expected to be easily realized in an artificial system.
Generally speaking, the moir\'{e} patterns imply the result of the
competition of at least two effects on electrons that are important both for
applications and for fundamental physics. These patterns can be the periodic
appearance of two different quantum phases, exhibiting unprecedented states
of matter. Two periodic structures can arise from periodic electric and
magnetic fields, respectively. The periodic potential (or optical lattice)
and the strong particle-particle repulsion can compactly array electrons
(atoms). The magnetic field with a slight mismatch period affects the spins
in two different ways, staggered manner or zero field, depending on the
location at the sample. It provides a simple way to demonstrate the moir\'{e}
patterns in a one-dimensional system which can be seen in Fig. \ref{fig1}.

In this paper, we study the moir\'{e} pattern that emerges in a dimerized
Kitaev spin chain with a periodic transverse field, which can be mapped onto
the system of dimerized spinless fermions with p-wave superconductivity. The
exact solution for staggered field demonstrates that there are two distinct
phases for the ground states. The strong field results in the Neel order of
spin array, while the spin liquid emerges for nonzero field due to the
isolated flat Bogoliubov--de Gennes band. When the period of the transverse
field slightly mismatches the lattice constant, the local properties, such
as the local staggered magnetization and local density of states, vary
periodically along the chain with a long beat period, indicating that two
phases, Neel and spin liquid, appear alternatively. We also propose a
dynamic scheme to detect the Moir\'{e} fringes experimentally. The
underlying mechanism is based on the relationship between the decay rate of
LE and the \textrm{LDOS} when a local perturbation is added. Numerical
simulation demonstrates that the decay rate exhibits the same periodic
behavior as the other quantities. It provides a method to detect the Moir\'{e%
} fringes in a photonic system.

This paper is organized as follows. In Section \ref{Model Hamiltonian}, we
present a dimerized Kitaev spin chain model with spatially modulated
transverse fields. In Section \ref{Magnetization, DOS and String correlation
function}, we introduce the concepts of Magnetization, DOS and string
correlation function to characterize the ground state properties. In Section %
\ref{Moire fringes}, we show the Moir\'{e} fringes in the model. In Section %
\ref{Dynamic detection}, we propose a dynamic scheme to detect the Moir\'{e}
fringes experimentally.\ Finally, we give a summary in Section \ref{Summary}%
.

\section{Model Hamiltonian}

\label{Model Hamiltonian}

We start our investigation by considering a dimerized Kitaev spin chain with
spatially modulated transverse fields
\begin{eqnarray}
H &=&\sum_{j=1}^{N}[(1-\delta )\sigma _{2j-1}^{x}\sigma _{2j}^{x}+(1+\delta
)\sigma _{2j}^{y}\sigma _{2j+1}^{y}]  \notag \\
&&+\sum_{j=1}^{2N}g_{j}\sigma _{j}^{z},  \label{H1}
\end{eqnarray}%
where $\sigma _{j}^{\alpha }$ ($\alpha =x,$ $y,$ $z$) are the Pauli
operators on site $j$ and $g_{j}=g\cos [\pi (1+\Delta )j]$. We take $\sigma
_{2N+1}^{y}\equiv 0$\ to impose the open boundary condition. In the
zero-field case $(g=0)$, it has been studied in many perspectives \cite%
{A.Kitaev,XYF,KLH}. In the staggered-field case $(\Delta =0)$, previous work
\cite{WR} devoted to the topological feature of the degeneracy lines. For an
arbitrary parameter distribution function $g_{j}$, we have an equivalent
dimerized spinless fermion model with the Hamiltonian%
\begin{eqnarray}
H &=&\sum_{j=1}^{N}[\left( 1-\delta \right) (c_{2j-1}^{\dagger
}c_{2j}^{\dagger }+c_{2j-1}^{\dagger }c_{2j})  \notag \\
&&+\left( 1+\delta \right) (c_{2j+1}^{\dagger }c_{2j}^{\dag }+c_{2j}^{\dag
}c_{2j+1})+\mathrm{H.c.}]  \notag \\
&&+g_{j}\sum_{j=1}^{2N}\cos [\pi (1+\Delta )j](1-2c_{j}^{\dagger }c_{j}),
\label{2}
\end{eqnarray}%
which describes the p-wave superconductivity. Here $c_{j}$ is spinless
fermionic operators and this mapping is obtained by the Jordan-Wigner
transformation \cite{P.Jordan}
\begin{eqnarray}
\sigma _{j}^{x} &=&-\prod\limits_{l<j}(1-2c_{l}^{\dagger
}c_{l})(c_{j}^{\dagger }+c_{j}),  \notag \\
\sigma _{j}^{y} &=&-i\prod\limits_{l<j}(1-2c_{l}^{\dagger
}c_{l})(c_{j}^{\dagger }-c_{j}),  \notag \\
\sigma _{j}^{z} &=&1-2c_{j}^{\dagger }c_{j}.
\end{eqnarray}%
The on-site external field $g_{j}$\ is extracted from the continuous field $%
g(x)=g\cos [\pi (1+\Delta )x]$ in the continuous coordinate $x$, with the
period $2/(1+\Delta )$. For sufficient small $\Delta $, we have $%
g_{j}\approx g_{\mathrm{eff}}\cos (\pi j)$, where $g_{\mathrm{eff}}=g\cos
(\pi \Delta j)$\ varies slowly. For a small scale, $g_{j}$\ can be regarded
as a staggered magnetic field with amplitude $\left\vert g_{\mathrm{eff}%
}\right\vert $. For a long scale, $\left\vert g_{\mathrm{eff}}\right\vert $\
is a periodic function of $j$\ with beat period $2/\Delta $. To investigated
the local properties of the system within a small region, we consider the
Hamiltonian as a homogeneous one, i.e., with zero $\Delta $\ but varied $g$.
The ground state properties with different $g$ reflect the local properties
of the original Hamiltonian within different space regions.

\begin{figure}[tbph]
\includegraphics[width=0.55\textwidth, clip]{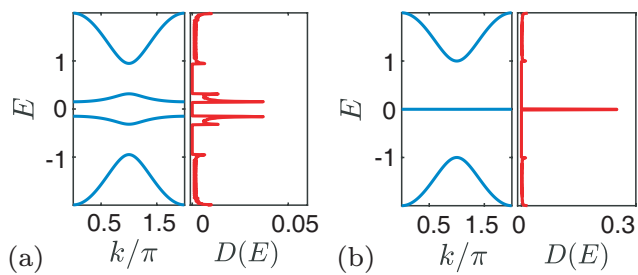}
\caption{Plots of the Bogoliubov-de Gennes energy bands for two typical
conditions (a) $g=0.4$, (b) $g=0.001$, as a function of momentum $k$ (blue
line) and $D(E)$ (red line), respectively. Notably, the flat band energy
value and the\ infinite \textrm{DOS} value both occur when $g\rightarrow 0$.
The size of the system is $N=500$, parameters are $\Delta =0$ and $\protect%
\delta =0.2$.}
\label{fig2}
\end{figure}

We impose periodic boundary condition $\sigma _{j}^{\alpha }\equiv \sigma
_{j+2N}^{\alpha }$ and perform the Fourier transformations for two
sub-lattices, which obeys
\begin{equation}
c_{j}=\frac{1}{\sqrt{N}}\sum_{k}e^{ikl}\left\{
\begin{array}{cc}
\alpha _{k}, & j=2l-1 \\
\beta _{k}, & j=2l%
\end{array}%
\right. ,
\end{equation}%
where $l=1,2,...,N$, $k=2m\pi /N$, $m=0,1,2,...,N-1$. Thus, spinless
fermionic operators in $k$ space $\alpha _{k},$\ $\beta _{k}$ can be
expressed as
\begin{equation}
\begin{array}{cc}
\alpha _{k}=\frac{1}{\sqrt{N}}\sum\limits_{l}e^{-ikl}c_{j}, & j=2l-1 \\
\beta _{k}=\frac{1}{\sqrt{N}}\sum\limits_{l}e^{-ikl}c_{j}, & j=2l%
\end{array}%
.
\end{equation}%
This transformation block diagonalizes the Hamiltonian with translational
symmetry, i.e.,%
\begin{equation}
H_{0}=\sum_{k}H_{k}=\sum_{k}\psi _{k}^{\dagger }h_{k}\psi _{k},
\end{equation}%
where
\begin{eqnarray}
H_{k} &=&\frac{1}{2}[\gamma _{k}(\alpha _{-k}^{\dagger }\beta _{-k}+\beta
_{-k}\alpha _{k}+\alpha _{-k}^{\dagger }\beta _{k}^{\dagger }+\beta
_{k}^{\dagger }\alpha _{k})+\mathrm{H.c.}]  \notag \\
&&-g(\alpha _{-k}\alpha _{-k}^{\dagger }-\alpha _{k}^{\dagger }\alpha
_{k}-\beta _{-k}\beta _{-k}^{\dagger }+\beta _{k}^{\dagger }\beta _{k})
\end{eqnarray}%
satisfies the relation $\left[ H_{k},H_{k^{\prime }}\right] =0$. Here, the
core matrix $h_{k}$\ obeys
\begin{equation}
h_{k}=\left(
\begin{array}{cccc}
0 & 0 & \gamma _{-k} & -g \\
0 & 0 & -g & 0 \\
\gamma _{k} & -g & 0 & 0 \\
-g & 0 & 0 & 0%
\end{array}%
\right) ,
\end{equation}%
\ \ which based on the basis vector
\begin{equation}
\psi _{k}=\frac{1}{\sqrt{2}}\left(
\begin{array}{c}
-\alpha _{-k}^{\dagger }+\alpha _{k} \\
-\beta _{-k}^{\dagger }+\beta _{k} \\
\beta _{-k}^{\dagger }+\beta _{k} \\
-\alpha _{-k}^{\dagger }-\alpha _{k}%
\end{array}%
\right) ,
\end{equation}%
where $\gamma _{k}=(1-\delta )+(1+\delta )e^{ik}$. The eigenvector with
eigenvalue can be solved as%
\begin{equation}
\varepsilon _{\rho \sigma }^{k}=\frac{\rho }{\sqrt{2}}\sqrt{\Lambda
_{k}+\sigma \sqrt{\Lambda _{k}^{2}-4g^{4}}}  \label{spc}
\end{equation}%
and%
\begin{equation}
|\phi _{\rho \sigma }^{k}\rangle =\frac{1}{\Omega _{\rho \sigma }}\left(
\begin{array}{c}
\varepsilon _{\rho \sigma }^{k}g\gamma _{-k} \\
\varepsilon _{\rho \sigma }^{k}[\left( \varepsilon _{\rho \bar{\sigma}%
}^{k}\right) ^{2}-g^{2}] \\
\text{ }g[\left( \varepsilon _{\rho \sigma }^{k}\right) ^{2}-g^{2}] \\
-g^{2}\gamma _{-k}%
\end{array}%
\right) ,
\end{equation}%
where parameters satisfy $\sigma ,\rho =\pm $ and $\Lambda _{k}=\left\vert
\gamma _{k}\right\vert ^{2}+2g^{2}$. The normalization factors are\textbf{\ }%
$\Omega _{\rho \sigma }=\rho \sqrt{2}g(\varepsilon _{\rho \sigma }^{k})^{-1}$%
\textbf{\ }$\{[(\varepsilon _{\rho \sigma }^{k})^{4}-g^{4}]$\textbf{\ }$%
[(\varepsilon _{\rho \sigma }^{k})^{2}-g^{2}]\}^{\frac{1}{2}}$\textbf{.}
There are four Bogoliubov-de Gennes\ bands from the eigenvalues of $h_{k}$,
indexed by $\rho ,\sigma =\pm $. In this study, we consider the case with
the band touching points for $\varepsilon _{+-}^{k}$ and $\varepsilon
_{--}^{k}$, which occur at $k_{c}$. With the equation
\begin{equation}
\Lambda _{k_{c}}=\sqrt{\Lambda _{k_{c}}^{2}-4g^{4}},
\end{equation}%
the solution of it obeys%
\begin{equation}
g=0,k_{c}\in (0,2\pi ],
\end{equation}%
which induce the flat zero band. These flat band can result in $N$-order
degree of degeneracy, which is a crucial motivation of this work. We plot
the energy structures for two typical values $g$ (Fig. \ref{fig2}).


\section{Magnetization, DOS and string correlation function}

\label{Magnetization, DOS and String correlation function}

For the uniform system $(\Delta =0)$ what we study here, the ground state is
strongly determined by the magnitude $g$. In fact, this viewpoint can be
seen from two limits.\textbf{\ }On the one hand,\textbf{\ }in strong limit $%
\left\vert g\right\vert \gg 1$, the ground state is a Neel magnet with
staggered spin alignment.\textbf{\ }On the other hand, the Hamiltonian has
been systematically studied \cite{KLH} when $g=0$.\textbf{\ }The ground
state is spin liquid phase, i.e., $A_{x}$\ or$\ A_{y}$\ phase for $\delta <0$%
\ or $\delta >0$, respectively.\ In the medium $g$, the crossover ground
state can be obtained from exact solution. In this Section, we utilize three
parameters, magnetization, \textrm{DOS} and string correlation function (%
\textrm{SCF}), to characterize the ground state properties.

\begin{figure}[tbph]
\includegraphics[ bb=5 16 471 388, width=0.3\textwidth, clip]{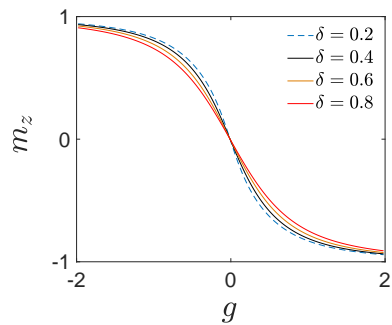}
\caption{Numerical simulations of $m_{z}$ as a function of on-site potential
$g$ for four typical values of $\protect\delta $: $\protect\delta =0.2$
(blue dashed line), $\protect\delta =0.4$ (black solid line), $\protect%
\delta =0.6$ (orange solid line) and $\protect\delta =0.8$ (red solid line).
The size of the system is $N=500$ and parameter is $\Delta =0$.}
\label{fig3}
\end{figure}

(i) The staggered magnetization on site $j$ is defined as%
\begin{equation}
m_{j}^{z}\equiv (-1)^{j}\left\langle G\right\vert \sigma _{j}^{z}\left\vert
G\right\rangle
\end{equation}%
for ground state $\left\vert G\right\rangle $. According to the
Hellmann-Feynman theorem, we have
\begin{equation}
m_{j}^{z}=(-1)^{j}\frac{1}{2N}\langle G|\frac{\partial H}{\partial g_{j}}%
|G\rangle =(-1)^{j}\frac{\partial E_{g}}{\partial g_{j}},  \label{mz}
\end{equation}%
where $E_{g}$\ is the many-particle density of ground state energy. For zero
$\Delta $, $m_{j}^{z}$\ is a uniform function of $(\delta ,g)$, i.e.,
\begin{equation}
m_{z}=m_{j}^{z}=\frac{\partial E_{g}}{\partial g},  \label{Mz}
\end{equation}%
With exact solution
\begin{eqnarray}
E_{g} &=&\frac{1}{4\pi }\int_{-\pi }^{\pi }\left( \varepsilon
_{-+}^{k}+\varepsilon _{--}^{k}\right) dk  \notag \\
&=&-\frac{2\sqrt{1+g^{2}}}{\pi }[\mathrm{E}(\frac{1-\delta ^{2}}{1+g^{2}})],
\end{eqnarray}%
in last Section, $m_{z}$\ can be exactly obtained as
\begin{equation}
m_{z}=-2g[\text{\textrm{E}}(\frac{1-\delta ^{2}}{1+g^{2}})]/(\pi \sqrt{%
1+g^{2}}).
\end{equation}%
Here, $\mathrm{E}(t)$ is the complete elliptic integral. Notably, parameter
satisfies $m_{z}=0$\ when $g=0$, which demonstrates one characteristic of
quantum spin liquid \cite{KLH}. More phenomena of the parameter $m_{z}$ are
described in Fig. \ref{fig3}.

(ii) On the other hand, the \textrm{DOS} for Bogoliubov-de Gennes\ band\ is
defined as%
\begin{equation}
D(E)=\lambda \frac{\mathrm{d}\mathcal{N}(E)}{\mathrm{d}E},  \label{DOS}
\end{equation}%
generally,\textbf{\ }$\mathrm{d}\mathcal{N}(E)$\textbf{\ }indicates that how
many energy levels are appeared in interval $[E,E+dE]$\ and the
normalization factors are\textbf{\ }$\lambda =\Delta E/4N$.\ Notably, $%
D(E)\rightarrow \infty $\ accords with the fact that flat bands occur at $g=0
$.\textbf{\ }Furthermore, Fig. \ref{fig2} demonstrates it.

(iii) The correlation length of quantum spin model can also demonstrate the
character of spin liquid phase. We introduce the \textrm{SCF} for the ground
state, which is defined as%
\begin{eqnarray}
&&O_{x}^{(2r+1)}(2j-1,2(j+r))  \notag \\
&=&\left\langle G\right\vert \sigma
_{2j-1}^{x}(\prod_{n=2j}^{2(j+r)-1}\sigma _{n}^{z})\sigma
_{2(j+r)}^{x}\left\vert G\right\rangle .
\end{eqnarray}%
\begin{figure}[tbph]
\includegraphics[ bb=5 16 471 388, width=0.3\textwidth, clip]{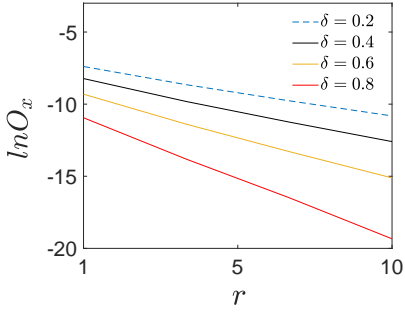}
\caption{Numerical simulations of a quantity $\ln O_{x}$ about correlation
length from Eq. (\protect\ref{CL1}) on uniform system, for four typical
classes of parameters $\protect\delta =0.2$ (blue dashed line), $\protect%
\delta =0.4$ (black solid line), $\protect\delta =0.6$ (orange solid line)
and $\protect\delta =0.8$ (red solid line), as a function of the long range
length $r$. The size of the system is $N=500$ and the parameter is $g=0.4$.
Note that the $\ln O_{x}$ decreases linearly with $r$ in finite $r$, which
are associated with the conclusion as given in Eq. (\protect\ref{CL0}).}
\label{fig4}
\end{figure}
For estimating the correlation length of \textrm{SCF}, we consider a
perturbated Hamiltonian
\begin{equation}
H^{(2r+1)}=H+c\sum\limits_{2j-1}\sigma
_{2j-1}^{x}(\prod_{n=2j}^{2(j+r)-1}\sigma _{n}^{z})\sigma _{2(j+r)}^{x}],
\label{H2}
\end{equation}%
where the extra term\ stands for long range spin-spin coupling with strength
$c$. The corresponding Schrodinger Equation for ground state is%
\begin{equation}
H^{(2r+1)}\left\vert G^{(2r+1)}\right\rangle =E_{g}^{(2r+1)}\left\vert
G^{(2r+1)}\right\rangle ,
\end{equation}%
where the\ many-particle density of ground state energy $E_{g,j}^{(2r+1)}$
can be exactly obtained as%
\begin{eqnarray}
E_{g}^{(2r+1)} &=&-\frac{1}{4\pi }\int_{-\pi }^{\pi }\{(1-\delta
)^{2}+(1+\delta )^{2}+c^{2}  \notag \\
&&+2(1-\delta )(1+\delta )\cos k+2c(1-\delta )  \notag \\
&&\times \cos (rk)+2c(1+\delta )\cos [(r+1)k]  \notag \\
&&+4g^{2}\}^{(1/2)}\mathrm{d}k
\end{eqnarray}%
\ by the translational symmetry of $H^{(2r+1)}$. Meanwhile, the
translational symmetry of $\left\vert G^{(2r+1)}\right\rangle $ results in%
\begin{eqnarray}
&&O_{x}^{(2r+1)}(2j-1,2(j+r))  \notag \\
&=&O_{x}^{(2r+1)}(2(j+l)-1,2(j+r+l))  \notag \\
&=&O_{x}^{(2r+1)}.
\end{eqnarray}%
According to the Hellmann-Feynman theorem, \textrm{SCF} can be expressed as%
\begin{eqnarray}
O_{x}^{(2r+1)} &=&\left[ \frac{\partial E_{g}^{(2r+1)}}{\partial c}\right]
_{c=0}  \notag \\
&=&-\int_{-\pi }^{\pi }\frac{1}{4\pi \sqrt{2}}\{(1-\delta )\cos
(rk)+(1+\delta )  \notag \\
&&\times \cos [(r+1)k]\}[(1+\delta ^{2})+(1-\delta ^{2})  \notag \\
&&\times \cos k+2g^{2}]^{-\frac{1}{2}}\mathrm{d}k.
\end{eqnarray}
The $O_{x}^{(2r+1)}$ satisfies the relation
\begin{equation}
O_{x}^{(2r+1)}\varpropto e^{-\frac{2r+1}{\xi }},  \label{CL0}
\end{equation}%
which can be transformed as%
\begin{equation}
\xi =2[\ln O_{x}^{(2r+1)}-\ln O_{x}^{(2r+3)}]^{-1},  \label{CL1}
\end{equation}%
Fig. \ref{fig4} numerically demonstrates it. Actually, $\xi $ is also
determined by $g$ and $\xi =0$ occurs at a large\ limit of $g$,\textbf{\ }we
also numerically \ demonstrate it (\ref{fig5}).

In a conclusion, this system has a large number of degeneracy ground states
with zero magnetization, which are the characterization of spin liquid
phase, as claimed in previous work \cite{KLH}. On the other hand, $D(E)=0$
vanishes, $\xi \rightarrow 0$ and $m_{z}\rightarrow 1$, when $g\rightarrow
\infty$, which all indicate the Neel phase.
\begin{figure}[tbph]
\includegraphics[ bb=5 0 471 388, width=0.3\textwidth, clip]{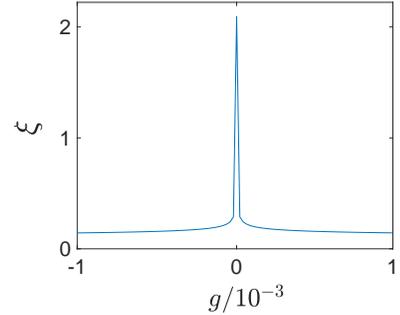}
\caption{Numerical simulations of correlation length $\protect\xi $ as a
function of on-site potential $g$. The correlation length $\protect\xi $ is
finite when $g\rightarrow 0$, while $\protect\xi $ decays rapidly to zero as
$\left\vert g\right\vert $\ increases. The size of system is $N=500$,
parameter is $\Delta =0$ and $\protect\delta =0.2$.}
\label{fig5}
\end{figure}

\section{moir\'{e} fringes}

\label{Moire fringes}

\begin{figure}[tbph]
\includegraphics[ bb=55 338 529 746, width=0.35\textwidth, clip]{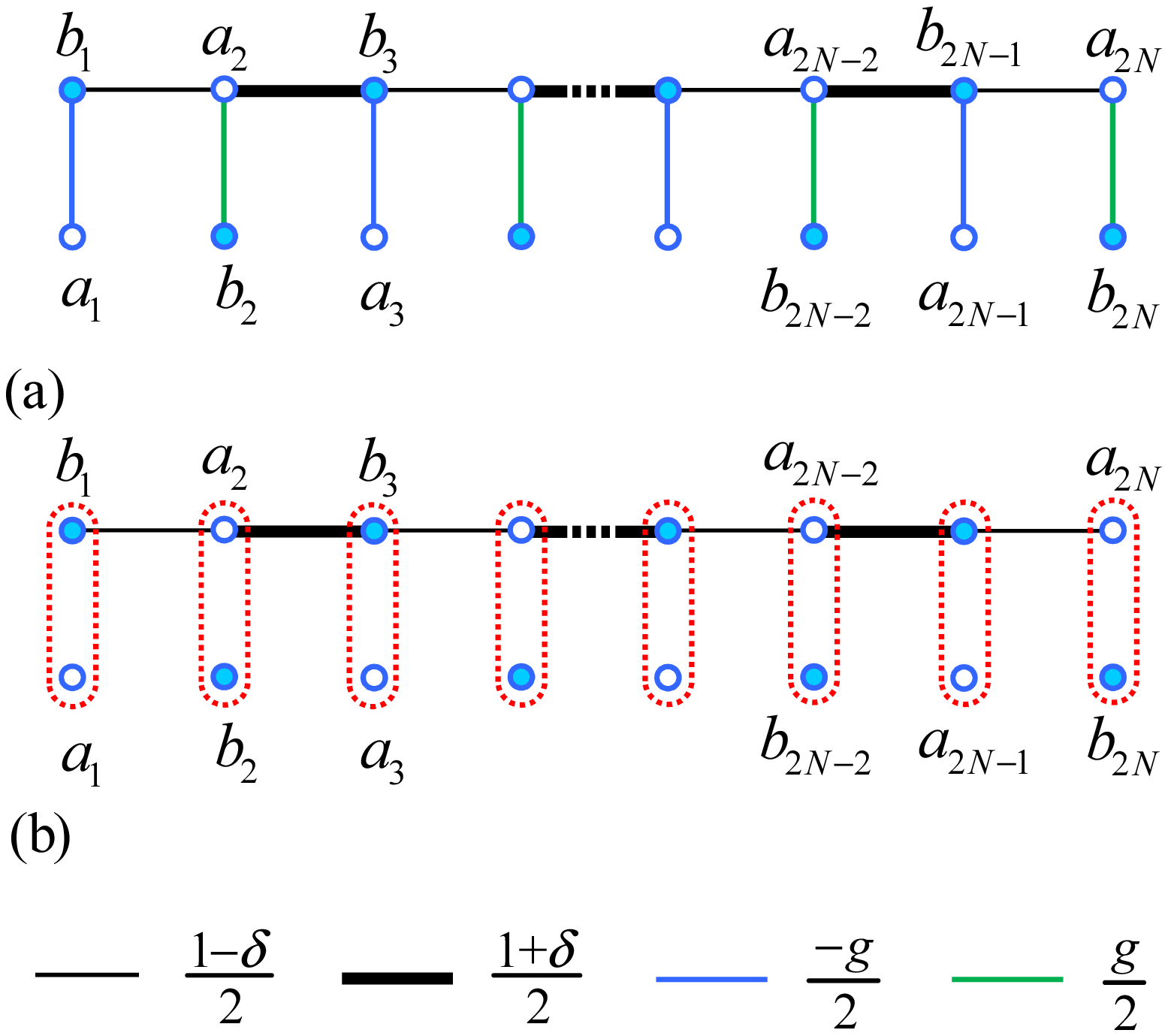}
\caption{Lattice geometries for the Majorana models, which is described in
Eq. (\protect\ref{H_M}). Solid (empty) circle indicates (anti) Majorana
modes. Panel (a) indicate a SSH chain with side-couplings ($1\pm \protect%
\delta )/2$ and $\pm g/2$, while panel (b) with $g=0$.}
\label{fig6}
\end{figure}

The aim of this work is to present the Moir\'{e} fringes of the system with
nonzero $\Delta $, properties of system for zero $\Delta $ but different $g$%
, discussed above, should appear in the different locations along the chain.
Consider magnetization $m_{j}^{z}$\ is a function of the coordinate, then we
introduce \textrm{LDOS} to replace \textrm{DOS}. For calculating such
parameters, the solution of $H$ is necessary. We utilize Majorana fermion
operators
\begin{equation}
a_{j}=c_{j}^{\dagger }+c_{j},b_{j}=-i(c_{j}^{\dagger }-c_{j}),  \label{ab}
\end{equation}%
which satisfy the relations
\begin{equation}
\left\{ a_{j},a_{j^{\prime }}\right\} =2\delta _{j,j^{\prime }},\left\{
b_{j},b_{j^{\prime }}\right\} =2\delta _{jj^{\prime }}
\end{equation}%
and%
\begin{equation}
\left\{ a_{j},b_{j^{\prime }}\right\} =0,a_{j}^{2}=b_{j}^{2}=1.
\end{equation}%
The inverse transformation obeys
\begin{equation}
c_{j}^{\dagger }=\frac{1}{2}(a_{j}+ib_{j}),c_{j}=\frac{1}{2}(a_{j}-ib_{j}).
\end{equation}%
Then the Majorana representation of Hamiltonian is%
\begin{equation}
H=\psi ^{T}h\psi ,
\end{equation}%
where $\psi ^{T}=(ia_{1},b_{1},ia_{2},b_{2},ia_{3},b_{3},...,ia_{2N},b_{2N})$%
.\ Here, $h$\ represents a $4N\times 4N$ matrix, which can be explicitly
wrote as%
\begin{eqnarray}
h &=&\frac{g}{2}\sum\limits_{j=1}^{2N}(-1)^{j}\cos (j\pi \Delta )\left\vert
a,j\right\rangle \left\langle b,j\right\vert   \notag \\
&&+\frac{1}{2}\sum\limits_{j=1}^{N}[(1+\delta )\left\vert a,2j\right\rangle
\left\langle b,2j+1\right\vert   \notag \\
&&+(1-\delta )\left\vert a,2j\right\rangle \left\langle b,2j-1\right\vert ]+%
\mathrm{H.c.},  \label{H_M}
\end{eqnarray}%
where $\left\vert \lambda ,j\right\rangle \ $is an orthonormal complete set,
which satisfies $\langle \lambda ,j\left\vert \lambda ,j^{\prime
}\right\rangle =\delta _{\lambda \lambda ^{\prime }}\delta _{jj^{\prime }}$,
$j\in \left[ 1,2N\right] $, $\lambda =a,b$ and $\left\langle
b,2N+1\right\vert \equiv 0$. Obviously, $h$\ describes a single-particle
tight-binding SSH chain with side-couplings and cosine-modulated couplings%
\textbf{. }The schematic of the Majorana lattice system is sketched in Fig. %
\ref{fig6}.\textbf{\ }

\begin{figure*}[tbph]
\includegraphics[width=1\textwidth, clip]{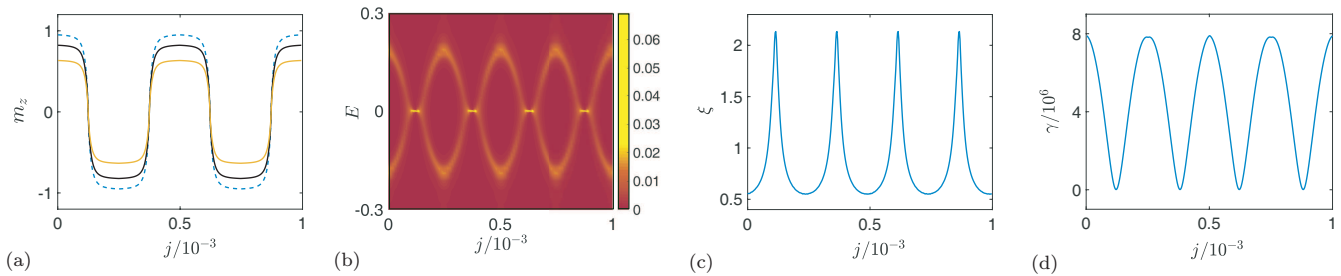}
\caption{Numerical simulations of four typical parameters: (a) magnetization
($m_{z}$), (b) local density of state (\textrm{LDOS}), (c) correlation
length ($\protect\xi $)\ and (d) decay of the \textrm{LE} ($\protect\gamma $%
), as a function of site $j$ respectively. Notably, these four panel are the
main results of this work. Panel (a), on-site potential amplitude parameters
are $g=-2.5$ (blue dashed line), $g=-1$ (black solid line) and $g=-0.5$
(orange solid line). The top (or valley) of the magnetization depend on the
value of $g$ in panel (a). All panels exhibit the unitary moir\'{e} patterns
for infinite site chain. Parameters in panel (d) is $\protect\eta =0.005$.
The size of system is $N=500$, the common parameters are $\Delta =0.002$, $%
g=0.4$ and $\protect\delta =0.2$.}
\label{fig7}
\end{figure*}

Then ground state energy $E_{g}$, $m_{j}^{z}$\ and $D_{j}$ (LDOS)\ can also
be obtained by diagonalizing the matrix of the lattice system. For the
ground state, Majorana operators $a_{2j-1}$\ and $b_{2j}$\ are free in $%
g_{j}=0$ region, leaving large numbers of degenerate ground states. We
depict the magnetization as a function of $j$ (Fig. \ref{fig7} (a)) and
compute $D_{j}$\ by using the method \cite{Haydock}(Fig. \ref{fig7} (b)).
Inspired by the characters of string correlation function, we consider an
artificial model with non-neighbour interactions, and the Hamiltonian obeys%
\begin{equation}
h_{j}^{(2r+1)}=h+\frac{c}{2}\left( \left\vert a,2(j+r)\right\rangle
\left\langle b,2j-1\right\vert +\mathrm{H.c.}\right) .
\end{equation}%
Density of ground state energy $E_{g,j}^{(2r+1)}$ for many-particle system
equals the sum of all negative eigenvalues of from $h_{j}^{(2r+1)}$ with the
coefficient $1/(2N)$, which can be numerically obtained. The corresponding
\textrm{SCF} and $\xi $\ can be computed from
\begin{equation}
O_{x,j}^{(2r+1)}=\lim_{c\rightarrow 0}\frac{\partial E_{g,j}^{(2r+1)}}{%
\partial c},
\end{equation}%
thus, we get the equation
\begin{equation}
\xi _{j}=2[\ln O_{x,j}^{(2r+1)}-\ln O_{x,j}^{(2r+3)}]^{-1},  \label{CL2}
\end{equation}%
and depict $\xi _{j}$ as a function of $j$\ (Fig. \ref{fig7} (c)).

\section{Dynamic detection}

\label{Dynamic detection}
\begin{figure}[tbph]
\includegraphics[ bb=5 12 471 388, width=0.3\textwidth, clip]{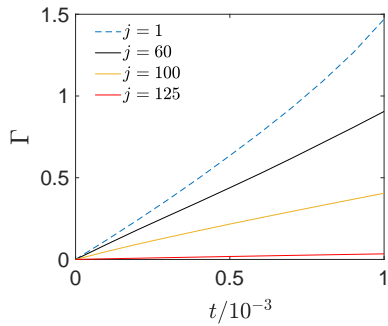}
\caption{ along the site $j$ from $1$ to $125$. The system parameters are $%
N=500$, $\Delta =0.002$, $\protect\eta =0.005$, $g=0.4$ and $\protect\delta %
=0.2$.}
\label{fig8}
\end{figure}
\begin{figure}[tbph]
\includegraphics[ bb=1 2 439 360, width=0.3\textwidth, clip]{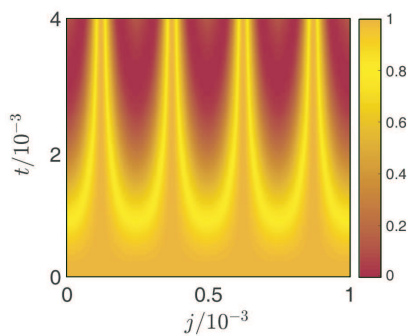}
\caption{Numerical simulation of \textrm{LE} $M(t)$ as a function of site $j$
and evolution time $t$ in Majorana model. The periodic behavior is obvious,
exhibiting the moir\'{e} pattern as expected. The size of system is $N=500$, the
parameters are $\Delta =0.002$, $\protect\delta =0.2$, $g=0.4$, and $\protect%
\eta =0.005$.}
\label{fig9}
\end{figure}
Motivated by the relationship between the decay rate of \textrm{LE} and the
\textrm{LDOS} when the local Hamiltonian is perturbed slightly, we want to
demonstrate the veiled information inside the model of this work. This
investigation may provide the scheme to\ experimentally detect the Moir\'{e}
fringes. It is a challenge to realize a quantum spin chain with the
Hamiltonian $H$. However, the Majorana lattice $h$ appears as a relative
simpler structure, which can be arranged in photonic system. it is based on
the analogy between light propagating through a photonic crystal and the
tight binding Hamiltonian. For instance, topological effects in some
electronic systems can be observed in their photonic counterpart \cite%
{MC,MC2,SS,JN,MH}. In a photonic platform, Pauli exclusion is not obeyed, a
single-particle state can be amplified by the large population of photons.
It allows for a high degree of control over the system parameters.

In this Section, we consider the Hamiltonian $h_{\eta }$ with a slight
perturbation, which has the form
\begin{equation}
h_{\eta }=h+\eta \sum\limits_{j=1}^{2N}(\left\vert b,j\right\rangle
\left\langle a,j\right\vert +\mathrm{H.c.}),  \label{Hn}
\end{equation}%
where the perturbation is a shift field with strength $\eta $. We employ the
\textrm{LE} to investigate the dynamical signature of the moir\'{e} pattern.
The \textrm{LE} for an initial state $|\psi _{0}\rangle $ is defined as%
\begin{equation}
M(t)=|\langle \psi _{0}|\exp (ih_{\eta }t)\exp (-iht)|\psi _{0}\rangle |^{2}.
\end{equation}%
We take the site-state $|\psi _{0}\rangle =\left\vert b,2j\right\rangle $\
as the initial state, thus, $M(t)$\ is a function of position, which labeled
by $M_{j}(t)$. Numerical simulations demonstrate the \textrm{LE} decays in
the form%
\begin{equation}
M_{j}(t)=e^{-\gamma _{j}t^{2}},  \label{M}
\end{equation}%
thus, we have the equation%
\begin{equation}
\Gamma _{j}(t)=\sqrt{-\ln M_{j}(t)}=\sqrt{\gamma _{j}}t.  \label{Gamma}
\end{equation}%
The parameter $\Gamma _{j}(t)$\ is introduced to present the decay
behaviour, as a function of evolution time $t$ (Fig. \ref{fig8}).
Furthermore, we depict the decay of \textrm{LE} ($\gamma _{j}$)\ as a
function of site $j$ (Fig. \ref{fig7} (d)). For clarity, we demonstrate the
relations of calculated \textrm{LE} for a serious $j$\ and $t$ values (Fig. %
\ref{fig9}).\ Notably, the performances of $\Gamma $, $\gamma $\ and $M$,
which are described in above three figures can not be affected by the varied
parameters $\delta $\ or $g$. It is worth mentioning briefly that such a moir%
\'{e} pattern can be implemented through two-dimensional array of
evanescently coupled optical waveguides. Additionally, the \textrm{LE} of
photons can be observed in a binary waveguide, by exchanging the two
sublattices after some propagation distance \cite{SLonghi}.

\section{Summary}

\label{Summary}

In summary, we have demonstrate a super periodicity in the coordinate space
along a quantum spin chain is imposed on it if the period of the external
sinusoidual magnetic field has a slight difference with the lattice
constants. There are two quantum phases in each period, one is Neel phase,
another is spin liquid phase. Additionally, we have proposed a dynamic
scheme to detect the Moir\'{e} fringes based on the measurement of \textrm{LE%
} with the respect to local perturbation. It provides a method to detect the
Moir\'{e} fringes in a photonic system, which is different from the real
quantum condensed-matter system, but can reproduce almost all
condensed-matter experiments.

\acknowledgments This work was supported by National Natural Science
Foundation of China (under Grant No. 11874225).

\end{document}